# Robotics for Learning

*By Dennis Toh, Ravintharan and Matthew Lim (Woodlands Ring Secondary School)*
*Wee Loo Kang, Matthew Ong (Educational Technology Division, MOE)*

## What

Teaching Robotics is about empowering students to create and configure robotics devices and program computers to nurture in students the skill sets necessary to play an active role in society.

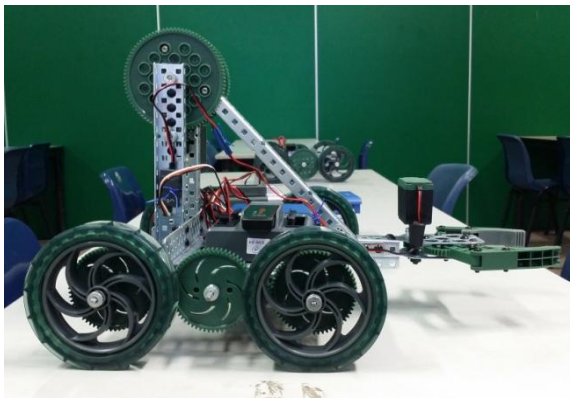

Figure 1. Example of a wheeled Vex robot. Photo from Woodlands Ring Secondary.

The robot in Figure 1 focuses on the design of scaffolds and physical assembly methods, coupled with a computer logic program to make that makes it move or behave in a very precise (remote controlled or autonomous) manner. This enables students to investigate, explore and refine the program to affect the robots.

**Technologies and Tools**
- Vex Robotics
- http://www.vexrobotics.com/
- LEGO WeDo
- LEGO NXT
- http://education.lego.com/en-us/preschool-and-school/lower-primary/7plus-education-wedo/

The Robotics approach takes into account the increasing popularity of Computer Science and the learning by doing (Schank, Berman, & Macpherson, 1999) approach to solve complex problems and use computers meaningfully in learning (Barron & Darling-Hammond, 2008; Jonassen, Howland, Marra, & Crismond, 2008). In Singapore, teachers and students in Woodlands Ring Secondary and Rulang Primary have incorporated robotics to varying extents into formal and informal curricula. In addition, other less expensive robotics tools can be used, depending on the curriculum or different requirements at international competitions.

*From idea to working prototype. Engage your students in practical, open-ended engineering challenges and problem solving.*

The two main pedagogical learning approaches that have taken root in Robotics are learning- by-doing and constructionism (Papert & Harel, 1991).

## How

**Formal Curriculum Infusion**

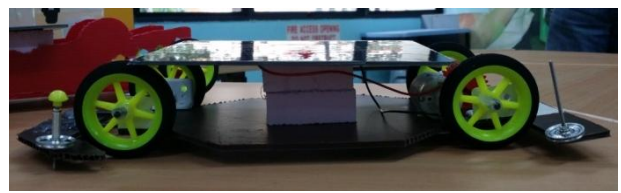

Figure 2. Example of a solar car built by students. Photo from Woodlands Ring Secondary.

Robotics education typically requires differentiated design and implementation to suit students with different learning needs and at different grade levels. In Rulang Primary, robotics was infused meaningfully into every primary level to varying extents across subjects such as English STELLAR resources, Maths, Science, Art and Music, in order to suit the syllabus. In Woodlands Ring Secondary, solar car building (Figure 2), the VEX Robotics education programme (Figure 3), and electronic projects give students opportunities to learn curriculum objectives by solving hands-on problems in Mathematics and Science (Figure 4), while building life skills like effective collaboration and teamwork.

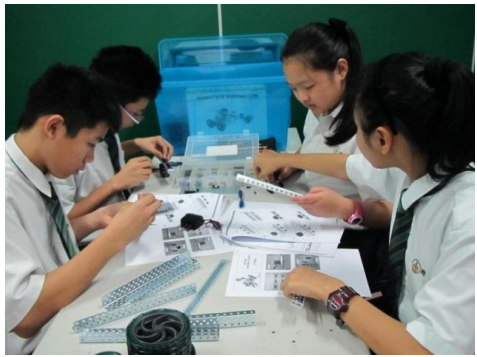

Figure 3. Students working on a 10 week 2-hourly programme to build robots under the Vex Robotics education programme. Photo from Woodlands Ring Secondary.

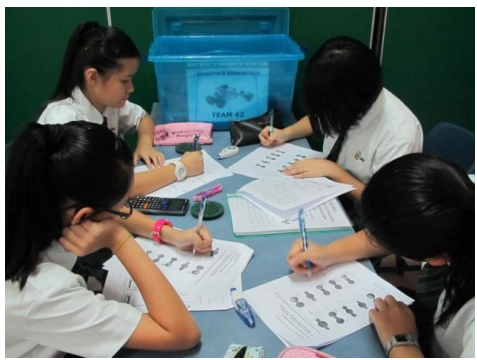

Figure 4. Students building the robot may be strategically engaged in formal curriculum learning through customised worksheets with applied Mathematics and Science. Students apply concepts learnt in Mathematics and Science to build better robots. Photo from Woodlands Ring Secondary.

**Informal Learning Opportunities**

Talent development programme can hone students' skills through opportunities to participate in national or international competitions. Such informal learning exposes students to more advanced programming, construction, building techniques, and can also improve students' presentation skills.

**Learning Framework**

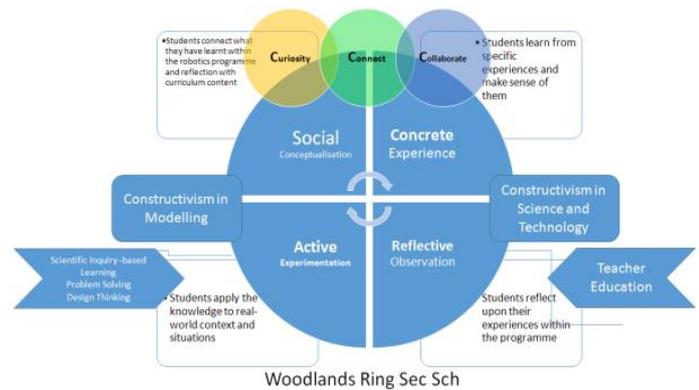

Figure 5. Woodlands Ring Secondary's Constructivism based Education-Learning Framework

A customised school-based learning framework (Figure 45) guides teachers, students and parents to better craft learning experiences (Wee, 2012). Examples are Woodlands Ring Secondary's Active-Social-Concrete-Reflective Cycle and Rulang Primary School's S.P.A.C.E. (Scenario, Perceiving the Problem, Asking Questions, Collaborative Learning and Evaluating).

**Partners**

Key partners in the Robotics programme include the Singapore Polytechnic Young Engineer's Club and schools with Robotics programme which create opportunities for students to enhance students' creative thinking as they overcome challenges.

**Schools**
- Woodlands Ring Sec
- Pei Hwa Sec
- Admiralty Secondary
- Chua Chu Kang Pri
- Compassvale Pri
- Rulang Primary
- Temasek Pri

In the next section, we take a look at how Woodlands Ring Secondary has designed an introductory robotics lesson to develop students' creativity and problem-solving skills.

**Lesson Plan**

**DURATION:** 120 minutes	**SUBJECT:**	Design and Technology

**LEVEL:** Sec 1 Express	**TOPIC:**	Applied Learning Programme – Robotics Education

*LEARNING OBJECTIVES*

*At end of the lesson package, pupils should be able to:*

- Develop basic process skills in problem-solving such as identifying the problem, comparing & contrasting ideas, discussing solutions, deciding on the best solution.

*MATERIALS AND RESOURCES REQUIRED FOR LESSON*

| ICT | Materials and Resources |
|---|---|
| VEX resources<br>PowerPoint slides<br>Visualiser | Forms 1-3 |

*LESSON PROCEDURES AND PEDAGOGY*

| Time | Procedures | Pedagogies Used | Materials & Resources |
|---|---|---|---|
| 20 mins | **Introduction**<br>• Ensure class is informed about their groupings one week earlier.<br>• Place team resources and files on the tables prior to start of lesson.<br>• Invite all students to be seated according to their designated groups and tables.<br>• Establish movement routines.<br>• Introduce the programme:<br>➢ Problem solving loop & Bloom's Taxonomy<br>➢ VEX standard set<br>➢ Sessions & Schedule<br>➢ Routines & Expectations<br>➢ The first 4 Thinking Skills | Collaborative learning | PowerPoint Slides |

| 45 mins | **Lesson Development**<br>Roles Assignment: Get each group to assign a Leader, Manager, Engineer & and Scribe among themselves.<br><br>Inform students that the skills they should acquire through the process are:<br>• Identifying the problem<br>• Comparing & contrasting ideas<br>• Discussing solutions<br>• Deciding on the best solution<br>Guiding questions:<br>• Why is there a need to do roles & responsibilities allocation?<br>• What are the advantages/disadvantages of solving problems as a group?<br><br>**Introduction to parts of the challenge**<br>Inform students there are 4 new thinking skills they should acquire through the game competition:<br>• Compare & Contrast and<br>• Organize & Classify (Toolbox)<br><br>These skills will be assessed through a given worksheet (Form 3)<br><br>Consolidate the learning by conducting a short discussion with the class. | Experiential learning | Form 1<br>(1 set/team)<br><br><br><br>PowerPoint Slides<br>Form 2<br>(1 set/team)<br><br><br><br>Form 3 |

| 25 mins | **Consolidation**<br>Teacher to demonstrate and allow students to experience handling the robotics tools:<br>- Handling of Allen Keys<br>- Screw, Kep Nut, Bearing Flat<br>- Collar & Axle<br><br>Note: In this segment, the teacher could focus on tightening skills<br><br>Allow students sufficient time to practice handling the tools on their own.<br><br>Teacher to monitor how students are handling the tools and provide necessary support where necessary. | Experiential learning | Visualiser |
|---|---|---|---|
| 30 mins | **Closure**<br>Recap that skills they should have learnt include:<br>- Identify, Discuss, Decide and Choose<br>- Compare & Contrast and Organise & Classify.<br>- Take team photos with names (optional) | Reflective learning | Papers, Camera |

# Robots Alive!

Robots are capable of amazing things. Some are able to fly like insects, clean homes, detonate bombs and even explore distant planets. At Woodlands Ring Secondary School, an ambitious plan has been put in place to prepare students to create some of their very own robots. Here, the focus is not on the robots but their creators – the students.

Robotics in Woodlands Ring Secondary School is about providing a highly engaging and practical learning experience to develop students' 21st century skills (Lemke, 2002). These include working in teams, being self-directed (Knowles, 1975) in their learning, and concerned about the environment.

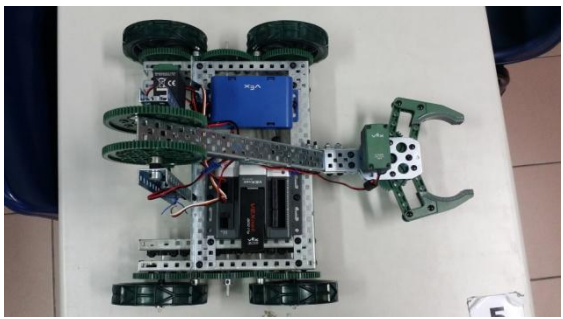

Figure 6.  Vex Robot built to pick up items and drop them at designated places via remote control. Photo from Woodlands Ring Secondary.

The journey begins in Secondary 1, where students learn problem-solving and critical thinking skills as they design a robot in teams. For 20 hours, students engage in practical tasks which include designing, building, programming and testing(Brown, 2008). For many students, it can prove to be delightfully challenging to build a moving vehicle from scratch to perform a series of elaborate manoeuvres, which include picking up items and dropping them in designated places. Through these tasks, students learn how to think systematically and analytically to solve a problem. Mr Matthew Lim, Senior Teacher in Robotics Education shared, "There's a lot of brainstorming, sketching, programming and iterations of designs. At times, the main challenge lies not in the problem-solving, but the problem-finding." Identifying the source of the problem where the robot needs to be fixed is as important, if not more, than finding a solution itself.

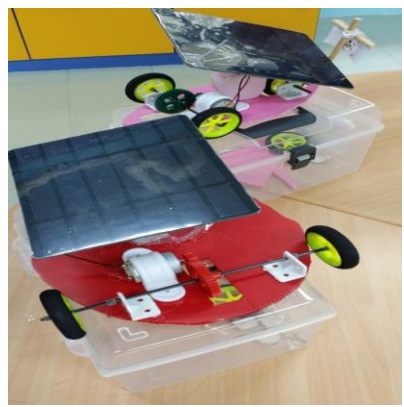

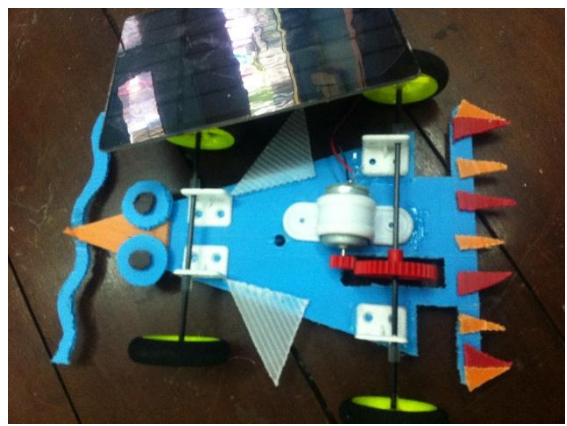

Figure 7.  Students' solar car with appropriate design for speed, gear ratio power, direction of travel and balance. Photo from Woodlands Ring Secondary.

The journey continues in Secondary 2, where Robotics is infused into the Design and Technology curriculum. Over the course of 30 hours, students design a solar energy-powered car and attempt to make it move as quickly as possible. This not only raises important technical questions about kinetic energy and friction but encourages students to think about larger global issues like alternative sources of clean energy. These teachable moments are peppered throughout the programme and students are provided multiple opportunities to transfer their learning to practical scenarios. One particular student noted that, "Some of the lessons can be applied in our daily lives, and also in Mathematics and Science."

At the end of the project, an exciting competition is organized where the fastest race cars from each class go head-to-head in an inter-class solar car racing competition. The top three winners from this competition are selected to represent Woodlands Ring Secondary School at the National Solar Prix Challenge. Needless to say, this motivates the students tremendously to do their best.

For students who are truly passionate about Robotics, the journey does not end in Secondary 2. In the third tier of the Robotics programme, students undertake a more elaborate electronics project in Secondary 3. This builds on their knowledge and skills from the previous projects and encourages students to appreciate electronics behind complex robots. They learnt to use multimeters, soldering and desoldering tools to construct sensor circuits. With these skills, students are able to design and build different functional robots, thereby discovering new knowledge through their creative imagination (Shanmugaratnam, 2002)**.**

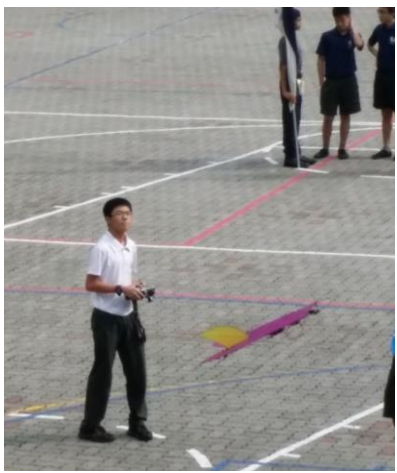

Figure 8. Student flying a remote-controlled aeroplane-motorized glider during co-curricular activity. Photo from Woodlands Ring Secondary.

On top of this extensive programme which is infused into the Secondary 1**, 2 and** 3 curricula, students are attracted to join one of the most sought after co-curricular activities in the school: The Robotics Club. In this club, students design remote-controlled robots that can move in multiple directions, lift objects, and crush soda cans. Some of these robots even include gravity-defying gliders that soar high above the school building and motion-sensing robots that play soccer with each other.

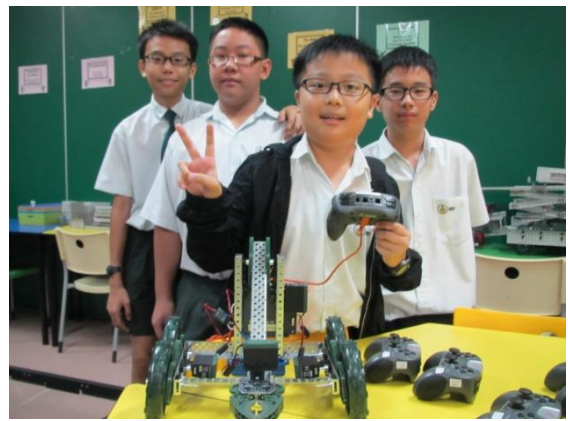

Figure 9. Toby (centre) and friends taking a victory photo after completing a Vex Robotics Programme. Photo from Woodlands Ring Secondary.

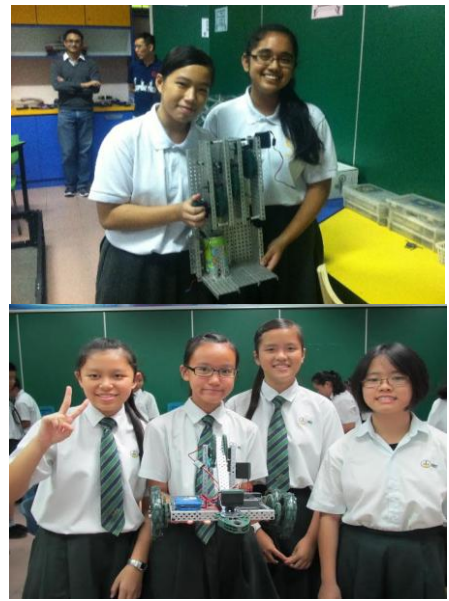

Figure 10. Enterprising students showcasing their talent and drive in their newly designed Vex robots. Photo from Woodlands Ring Secondary.

Despite all the inspiring creations, Woodlands Ring Secondary School is careful to keep the focus on the creator. Whether learners are exploring Robotics within the curriculum or their CCA, the key focus is on developing critical thinking, collaborative skills, and concern about the environment. Toby, a Secondary 1 student, summed it up nicely when he said, "Robotics enables you to learn life skills (Plomp, 2011)."